\documentclass[twocolumn,showpacs,showkeys,amsfonts,amssymb]{revtex4}

\usepackage{amsmath}

\usepackage[dvips]{graphics,graphicx,color}
\usepackage{epsfig}
\usepackage{subfig}

\begin{document}

\preprint{}

\title{Universality of Cluster Dynamics}

\author{Carson McFadden}
\email[]{cmcfad@chem.ucla.edu}
\affiliation{Department of Chemistry and Biochemistry, University of California, 607 Charles E. Young Dr. East, Los Angeles, CA 90095}

\author{Louis-S. Bouchard}
\email[]{bouchard@chem.ucla.edu}
\affiliation{Department of Chemistry and Biochemistry, California NanoSystems Institute, Biomedical Engineering IDP, and Jonsson Comprehensive Cancer Center, University of California, 607 Charles E. Young Dr. East, Los Angeles, CA 90095}

\date{\today}

\begin{abstract}
We have studied the kinetics of cluster formation for dynamical systems of dimensions up to $n=8$ interacting through elastic collisions or coalescence.  These systems could serve as possible models for gas kinetics, polymerization and self-assembly.  In the case of elastic collisions, we found that the cluster size probability distribution undergoes a phase transition at a critical time which can be predicted from the average time between collisions.   This enables forecasting of rare events based on limited statistical sampling of the collision dynamics over short time windows.  The analysis was extended to L$^p$-normed spaces ($p=1,\dots,\infty$) to allow for some amount of interpenetration or volume exclusion.  The results for the elastic collisions are consistent with previously published low-dimensional results in that a power law is observed for the empirical cluster size distribution at the critical time.  We found that the same power law also exists for all dimensions $n=2,\dots,8$, 2D L$^p$ norms, and even for coalescing  collisions in 2D.  This broad universality in behavior may be indicative of a more fundamental process governing the growth of clusters.  
\end{abstract}


\maketitle

\section{Introduction}

This paper is a study of the statistical behavior of the dynamics of clusters which are allowed to interact through elastic collisions or by coalescence.  The elastic collision dynamics are based on a ballistic billiard model analyzed theoretically by Sinai~\cite{bib:sinai,bib:sinai2}.  Cluster growth and self-assembly processes are relevant to a variety of research fields, including chemistry, materials science, physics and earth sciences. The study of such random processes can reveal information on the nature of collective interactions as well as make predictions on the occurrence of  rare and catastrophic events.  Early theoretical studies on cluster dynamics originate in the work of Bogoliubov, who showed that in the gas phase, groups of particles with short-ranged interactions behave like independent clusters~\cite{bib:bogolyubov}. Sinai provided a proof of cluster dynamics for colliding billiards for one dimensional (1D) systems~\cite{bib:sinai}, and subsequently for higher dimensions (restricted to sufficiently low densities). Sinai also proved ergodicity of the classical billiard model~\cite{bib:sinai2}.  The statistical properties of cluster dynamics has been studied for the 2D case with frictionless elastic billiards~\cite{gabriel}. In this paper we extended this statistical analysis of ballistic billiards to higher dimensions ($n$) up to $n=8$, higher densities ($\rho$), L$^p$-normed distance metrics ($p=1,\dots,\infty$) and to the case of coalescing billiards. We have found a high degree of universality which suggests that the dynamics of clustering are relatively independent of the details.

In an ensemble of interacting particles we may observe a phase transition where a dominant cluster emerges~\cite{gabriel}. In a classical Sinai billiard consisting of elastic collisions, the phase transition in the empirical density of clusters is not necessarily associated with a phase transition of the physical system in the traditional sense, such as a transition from liquid to solid or gas to liquid as function of temperature or pressure.  Instead, one observes a change in the empirical cluster density -- which plays the role of the order parameter -- as function of time.  Thus, it is indicative of the dynamics of the motion rather than a configurational change resulting from the variation of an intensive variable.  The collisions between billiards represent the interactions between parts of a system, and the transition to a dominant cluster that emerges is a manifestation of the events leading to a major catastrophic event. In the case of coalescence, the phase transition in the probability density can be associated with a physical change in the properties of the system.   Recent examples of the analysis of phase transitions in probability densities include earthquake prediction~\cite{keilis, gabriel2, gabriel3, zal3, zal4}, economic modeling~\cite{keilis2} and models of river networks~\cite{zaliapin}.   Established premonitory patterns have allowed the modeling of events in complex systems to be predicted using observed background activity ~\cite{gab2}. The prospect of predicting or controlling critical events occurring in a dynamic and complex environment is of broad interest. 

In the first part of the paper, we expand the study of dynamical phase transitions to higher dimensional elastic billiards and describe the statistics of the collisions in $n$-dimensional L$^p$-normed spaces, with $n=1,\dots,8$ and $p=1,\dots,\infty$.  The first result which emerges is the existence of the phase transition in dimensions greater than $n$=2, for higher densities, and for different L$^p$-normed spaces ($p=1,\dots,\infty$).  In the Euclidean norm case of Sinai billiards, the critical time appears to be independent of the dimensionality of the system ($n$).  Instead, this critical time solely depends on the average time between collisions $\langle \tau_n \rangle$.  Another notable finding is that the empirical cluster distribution at the critical point obeys a power law across all dimensions, densities and norms with the same exponent. 

In the second part of the paper, we allow the billiards to coalesce and form larger clusters.  These collisions can be analyzed using a binary tree model~\cite{zaliapin} first developed to analyze environmental transport in river networks.  Coalescence and coagulation are phenomenon that are present in many areas of chemistry~\cite{kang, ghosh}. Theories of coalescence date back to the work of Smoluchowski~\cite{smoluchowski,smoluchowski2} in the early 20th century, establishing the evolution of the concentration, $c_k(t)$, of clusters of mass $k$  using a master equation of the form (discrete case):
\begin{multline}
\frac{dc_k(t)}{dt} = \frac{1}{2} \sum_{i+j=k} K_{ij}c_i(t)c_j(t) - c_k(t)\sum_{j>0}K_{jk}c_j(t)
\end{multline}
where $K_{ij}$ is the interaction kernel, which is dependent on the collision process of $i$-mers and $j$-mers. The first term predicts an increase in $c_k(t)$ due to coalescence of an $i$-mer and $j$-mer; the second term deals with the decrease in $c_k(t)$ due to $k$-mers coalescing with clusters of different sizes ~\cite{kang}. The theory is based on two important assumptions: coalescence upon collision and the absence of hydrodynamic interaction between the different $i$-mers~\cite{wang}. Recent work in the field has led to corrections to Smoluchowski's equation. Such works include film drainage theory~\cite{chesters} and studies considering a hydrodynamic interaction term~\cite{wang, zeichner}.

The coalescence process we analyze is similar to that of Smoluchowski in the sense that there are no interparticle interactions except for coalescence events which occur upon contact, and the process begins at $t=0$ with a monodisperse collection of monomers. The model is found to exhibit similar properties with regards to universality of the phase transition as the elastic model. The results suggest that coalescing processes are governed by principles similar to that of non-coalescing billiards.

\section{Model and Definitions}

We start with a Sinai billiard~\cite{gabriel} in $n$ dimensions, involving $N$ spheres positioned inside a frictionless hyper-cubic domain.  The total mass, $m$, of the billiards in each case is \verb+1.0+ with radius $R$. The domain is the set of points

\begin{multline}
\{\Lambda = (x_{1}, x_{2}, \dots, x_{n}) : \\ 0<x_{1}<1; 0<x_{2}<1; \dots ; 0<x_{n}<1\}
\end{multline}

\noindent The density of billiards within this domain is:

\begin{eqnarray}
\rho = \frac {V_n(R)N}{V_{\Lambda_{n}}}
\end{eqnarray}

\noindent where $V_{n}(R)$ is the volume of a $n$-dimensional hypersphere of radius $R$ and $V_{\Lambda_{n}}$ is the volume of the hyper-cubic domain.

Clusters are defined in the elastic collision model using the notion of a $\Delta$-cluster~\cite{sinai2}. Time is measured by the variable $t$, and $\Delta$ specifies an interval of time.  A $\Delta$-cluster is a group of billiards which have effected each others' kinematics in the previous time interval $\Delta$. A $\Delta$-neighbor is defined as two billiards at time $t$ which have collided during the time interval $[t-\Delta, t]$. The set of all billiards which have interacted and are linked by $\Delta$-neighbor relationships, are called a $\Delta$-cluster.  A $\Delta$-cluster's mass, $M$, is the sum of the mass of the billiards which make up the cluster. We use the notation $M^i_{\Delta}(t)$ at time $t$ to represent the mass of the $i^{th}$ largest $\Delta$-cluster (by mass) and $N_{\Delta}(t)$ as the total number of clusters at the time $t$.   

For the coalescing case, clusters are defined as $\Delta$-neighbors or $\Delta$-clusters only if they coalesce upon collision. Each cluster corresponds to one billiard in the hypercubic domain.  $M^i_{\Delta}(t)$ is the mass of the $i^{th}$ largest $\Delta$-cluster at time $t$ and $N_{\Delta}(t)$ is the total number of clusters at time $t$.

\subsection{Kinetics of Collisions}

We use a ballistic colliding billiard model for both the elastic and coalescing cases.  We use an \textit{a priori} method to detect collisions in the Euclidean norm case and coalescing case. When varying the $p$ norm ($p \ne 2$), we use an \textit{a posteriori} method to detect collisions. Each of these are further described below.

\subsubsection{Sinai Billiard}

 Elastic hard sphere collisions are enforced for the Sinai billiard. Total energy, $E$, and momentum, $m\mathbf{v}$, of the billiards in the system remain constant:

\begin{align}
E = \sum_{i=1}^N\frac{m |\mathbf{v}_i|^2}{2}, \qquad m\mathbf{v} = \sum_{i=1}^N m\mathbf{v}_{i}
\end{align}

\noindent where $|\mathbf{v}_{i}|^2=(v^i_{1})^2 + ... + (v^i_{n})^2$.   Incident and reflective angles are identical for billiard-wall collisions.

In constructing the simulation, an \textit{a priori} method was used to calculate the next time, $t_{next}$, of collision between two billiards or between billiard and wall. The system was advanced to this time, $t_{current}=t_{next}$, the velocities redefined for colliding billiards, clusters recorded, and the next collision time, $t_{next}$, computed. 

\begin{figure}[t]
\includegraphics[scale=0.65]{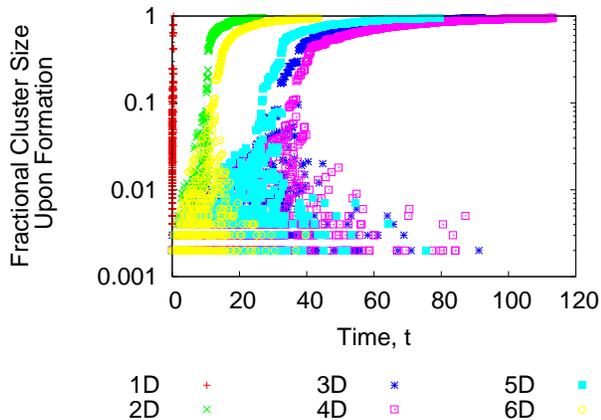}
\caption{\label{fig:fig1} Fractional cluster masses $M^i_{\Delta}$ as a function of time when they were formed for $N = 10^3$, $\rho = 0.001$, in dimensions as indicated. Each point corresponds to the creation of single cluster of fractional cluster mass $M^i_{\Delta}$ at time $t$.  A phase transition is observed where a dominant cluster emerges at a particular critical time $t_{c,n}$, which is different for each $n$ dimension. Typical results are presented for each dimension model, each with equal densities and number of billiards.  }
\end{figure}

\subsubsection{L$^p$-Normed Spaces}

We also investigated the effects of using an L$^p$-norm ($p \ne 2$).  The motivation for this is to consider particles that are of irregular shape and surfaces not unique to Euclidean space in collisions.  For the application of L$^p$-norms in this context, see~\cite{gamba}. The L$^p$ norm is

\begin{eqnarray}
\parallel x \parallel_p = ( |x_1|^p + |x_2|^p + ... + |x_n|^p )^{1/p}
\end{eqnarray}

\noindent  When $n$-dimensional billiards collide in any L$^p$-norm, momentum and energy are conserved by reassigning velocities as follows:

\begin{enumerate}
\item The vector normal to the sphere's surface at the point of collision is calculated between two colliding billiards, $a$ and $b$. First, by calculating the normal component along each dimension, $\vec{\eta} = (\eta_1, \eta_2, \dots,\eta_n)$ using the billiard positions $\vec{r_a} = (r_{a1}, \dots,r_{an})$ and $\vec{r_b} =(r_{b1}, \dots,r_{bn})$. 

\begin{align}
\eta_i =& \arrowvert p \cdot (r_{ai} - r_{bi})^{p-1} \arrowvert
\end{align}

where $1 \le i \le n$. We then calculate the associated unit normal vector $\hat{\eta}$:

\begin{align}
  \Arrowvert \vec{\eta} \Arrowvert^2 =& \sum_{k=1}^{k=n} \eta_k^2, \qquad  \hat{\eta} = \frac{\vec{\eta}}{\Arrowvert \vec{\eta} \Arrowvert }
\end{align}

\item At the time of collision, the initial (i) velocity of $a$ is given by  $\vec{v}^i_a = ( v^i_{a1}, \dots,  v^i_{an} )$. We denote the velocity for $b$ analogously. The initial relative velocity of $a$ and $b$ is calculated and dotted with the unit vector, $\hat{\eta}$, to find the speed, $v^i_{r}$ associated with the impulse.

\begin{eqnarray}
v^i_{r} = \hat{\eta} \cdot (\vec{v}^i_{a} - \vec{v}^i_{b}) 
\end{eqnarray}

\item The impulse, $\vec{I}= ( I_{1}, \dots, I_{n} ) $, is

\begin{eqnarray}
\vec{I} = 2  \frac{m_a m_b}{m_a + m_b}v^i_{r} \hat{\eta}
\end{eqnarray}

\item Velocities are reassigned for the two billiards to find the final velocity (f) after collision.

\begin{equation}
\vec{v}^f_{aj} = \vec{v}^i_{aj} - \vec{I}_j / m_a, \qquad \vec{v}^f_{bj} = \vec{v}^i_{bj} + \vec{I}_j / m_b
\end{equation}

\end{enumerate}

\noindent In implementing the L$^p$ norm dynamics, an \textit{a posteriori} method is used where a constant time interval, $\delta t$, was used to advance the billiards at each time step. Billiard-billiard intersections or billiard-boundary overlaps are used to identify collisions. When those collisions are identified, velocities are redefined using the method described above. The model then proceeds to the next time step.  The model was verified against the \textit{a priori} model in the $L^2$ (Euclidean) case in order to find a time step that is appropriate for simulation speed.

\begin{figure}[t]
\includegraphics[scale=0.7]{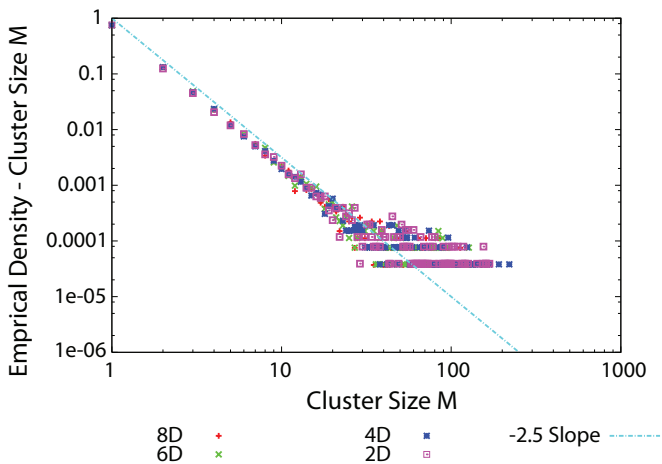}
\caption{\label{fig:fig2}  The power-law empirical cluster size distribution at the critical time, $t_{c}$ for $N=1000$ billiards at a density of $\rho = 10^{-4}$, for the different dimensions as indicated. One can see the approximate power law relationship, with $\beta \approx 5/2$. The magnitude of the cluster size, $M$, is plotted along the horizontal axis; the empirical cluster density of cluster size $M$ is given on the vertical axis based on averaging from 50 trials.  }
\end{figure}

\subsubsection{Coalescing Billiard}

Coalescing billiards were introduced to model polymerization reactions and self-assembly processes.  In a different but mathematically similar context, Zaliapin~\cite{zaliapin} studied transport in river networks.  This description of coalescing processes can include, for example, emulsions of oil in water, or colloidal particles flocculation in percolation analysis~\cite{hasmy}. As a motivation for the analysis one may consider a growing spherical polymer or colloid particle. The probability of coalescence for collisions of particles is assumed to follow an Arrhenius law:

\begin{eqnarray}
k = \exp \left( -\frac{E_{a}}{T_k} \right)
\end{eqnarray}

\noindent where $T_k$ is a temperature and $E_{a}$ is the activation energy required for coalescence.  For individual collisions $k$ is the conditional probability of a single coalescence event involving two billiards in a collision taking place where $E_{a}$ is fixed and $T_k$ is the total kinetic energy of the two colliding billiards.  Comparing a uniformly distributed random number in the interval $\{ z | 0 \le z \le 1 \}$ to $k$,  coalescence of the two $n$-spheres proceeds if $z < k$; otherwise, an elastic collision occurs.  In the latter case, the kinematics of the elastic collision are the same as that described for the Sinai billiards. In the former case (event of coalescence), the two coalescing billiards, designated \textit{daughters} $a$ and $b$, join to form one billiard, the \textit{parent, $\pi$}. In doing so, the radius, $R_{\pi}$, of the parent is defined in terms of the volumes of both initial clusters as to maintain constant density of billiards in the domain of the $n$-dimensional system,

\begin{eqnarray}
R_{\pi} = \left[ (R_{a})^n + (R_{b})^n \right]^{1/n}.
\end{eqnarray}

\noindent Mass is also conserved, with $m_\pi = m_a + m_b$, where $m_a$ and $m_b$ are the masses of daughters $a$ and $b$ respectively. The resulting magnitude of velocity of the parent billiard is defined by conservation of energy:

\begin{eqnarray}
\frac{{m_{\pi} \mathbf{v}_{\pi}}^2}{2} = \frac{m_{a} \mathbf{v}_{a}^2}{2} +\frac{m_{b} \mathbf{v}_{b}^2}{2}.
\end{eqnarray}

\noindent The direction of the parent's velocity vector is defined as that which results from a completely inelastic collision between the two daughter billiards where momentum is conserved. The parent's center, $(x_{1}, x_{2}, \dots, x_{n})$, is defined as the center of mass of the two daughter billiards. If the parent extends beyond the boundary of the system upon definition, its center is redefined perpendicular to the boundary edge so that it is completely within the confines of the system. In the event that the parent overlaps with another billiard in the system, that billiard undergoes a $k=$1.0 probability collision with the parent in the same manner as described above.  

As time evolves and billiards collide and coalesce, we obtain a binary tree of coalesced billiards. Each cluster is an individual growing billiard on the surface.  The total number of clusters is designated as $N_{A}(t)$ at time $t$ and $M^i_{A}(t)$ represents the mass of the $i^{th}$ largest cluster.  Due to the computationally demanding nature of this simulation, we investigated only the Euclidean distance metric (L$^2$) together with the \textit{a priori} method.

\subsection{Model Parameters}

Both models were simulated varying $N$, the number of billiards $100\le N\le 5 \cdot 10^3$ ; $\rho$, the density $10^{-6}\le\rho\le10^{-1}$; $E_{a}$, the activation energy satisfies $0\le E_a$. At time $t=0$, non-overlapping billiards are randomly placed in the volume 

\begin{multline}
\tau' = \{ (x_{1}, x_{2}, \dots, x_{n}) :  R<x_1<1-R; \\
\dots ; R<x_{n}<1-R \}
\end{multline}

\noindent  Particles are assigned an initial Maxwellian distribution:

\begin{eqnarray}
f( \mathbf{v} ) = \left( \frac{m}{2\pi T} \right)^{n/2} e^{- \frac{m| \mathbf{v} |^2}{2 T} }.
\end{eqnarray}

\noindent Several L$^p$ norms in the range $1 \le p \le  \infty$ were investigated. The temperature is held constant at $T = 1$.

\subsection{Phase Transition in Cluster Dynamics}

During an interval $0\le t \le t_{c}$ the cluster distribution evolves continuously and without gaps.  In this regime, the size of the largest cluster at any time, $M^1_\Delta(t)$, is not substantially larger than that of the second largest cluster, $M^2_\Delta(t)$. A dramatic change (Figure ~\ref{fig:fig1}) occurs at $t=t_{c}$ typical of a phase transition in which a dominant cluster appears.  The following definition of the critical time $t_c$ has been proposed~\cite{gabriel}

\begin{eqnarray}
t_c = \inf \{ t : M^1_t > M^i_r, r > t, i > 1 \}
\end{eqnarray}

\noindent This definition has the operational disadvantage of not being a stopping time, meaning that the occurrence of the phase transition cannot be decided based on previous knowledge of history available until the present time.  Our results suggest that despite the lack of a stopping time definition, perhaps an even more important observation is the empirical dependence of $t_c$ on the average time between collisions, $\langle \tau_n \rangle$, highlighting a potential way to predict its occurrence given some amount of statistical sampling.  The prediction of $t_c$ is important for the analysis of earthquake events~\cite{keilis, gabriel2, gabriel3, zal3, zal4}, interacting billiards ~\cite{sinai}, economic models~\cite{keilis2} and geographical river models~\cite{zaliapin}. In each of these instances, it has been suggested that cluster dynamics could be used to further understand the system or identify and predict the occurrence of a critical event.

\section{Results} 
\begin{figure}[t]
\includegraphics[scale=0.72]{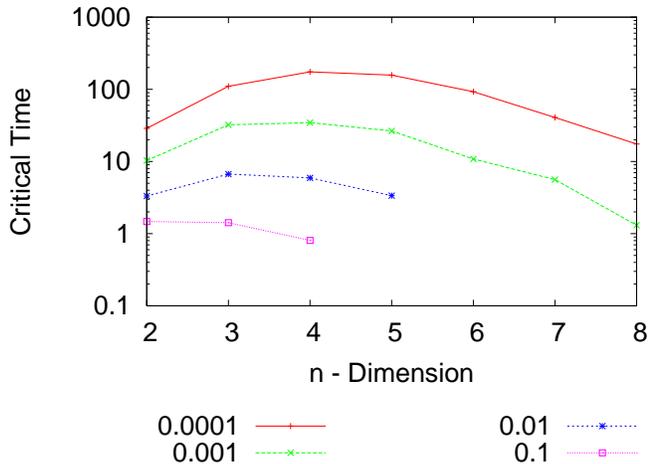}
\caption{\label{fig:fig3} Average critical times, $t_c$, plotted for each dimension with $N=1000$ varying density as indicated.}
\end{figure}

\subsection{Onset of the Phase Transition} 

In Fig.~\ref{fig:fig1} we vary the dimension with initial parameters $\rho = 10^{-4}$, $N=1000$. For each case, $n$-D,  a similar characteristic formation of clusters occurs with respect to time. The 2D case is in agreement with results previously published by Gabrielov {\it et al.}~\cite{gabriel}.  For each $n$ a phase transition is found at a critical time, $t_{c}(n)$, associated with a rapid increase in the growth of a dominant cluster, $M^1_{\Delta}(t_c)$.  

As the dimension $n$ is varied, while keeping other parameters constant, the time scale for the evolution of cluster dynamics is altered, as seen in Fig.~\ref{fig:fig1}. There are two contributions leading to this change.

\begin{enumerate}
\item High packing densities become less available for packing spheres at higher dimensions.  This point is discussed by Skoge {\it et al.}~\cite{skoge}. 
\item The critical time, $t_c$, is found to be dependent only on the average time between billiard collisions, $\langle \tau_n \rangle$ for $n \ge$ 2. This empirical dependence is shown in Fig.~\ref{fig:fig3} for approximately equal $\langle \tau_n \rangle$ at constant $N$.  
\end{enumerate}

The average time between collisions, $\langle \tau_n \rangle$  is calculated from

\begin{equation}
\langle \tau_n \rangle = \frac{t(M_{\Delta}^1 \approx 0.95)}{Z(\Delta)}
\label{eq:17}
\end{equation}

\noindent where $Z(\Delta)$ is the total number of billiard-billiard collisions that have taken place  in the time interval $[0, \Delta]$.

For fixed density $\rho$ the average time between collisions $\langle \tau_n \rangle$ first increases, then decreases with increasing $n$, as explained by the first point. Due to the critical time's dependence on $\langle \tau_n \rangle$, it scales accordingly. There is a general trend that the dimension, $n$, for which $\langle \tau_n \rangle$ reaches its maximum value, increases with decreasing density. Thus, we observe similar cluster distributions at each dimension, scaled by a factor.  From our knowledge of the behavior of systems of constant density and varying dimension, we have that as the density $\rho$ decreases, the largest critical time is associated with higher dimensions $n$.

\begin{figure*}[t]
\subfloat[$t_c$ as function of $\langle \tau_n \rangle$][]{\includegraphics[trim = 0 0 0 0, scale=1.0]{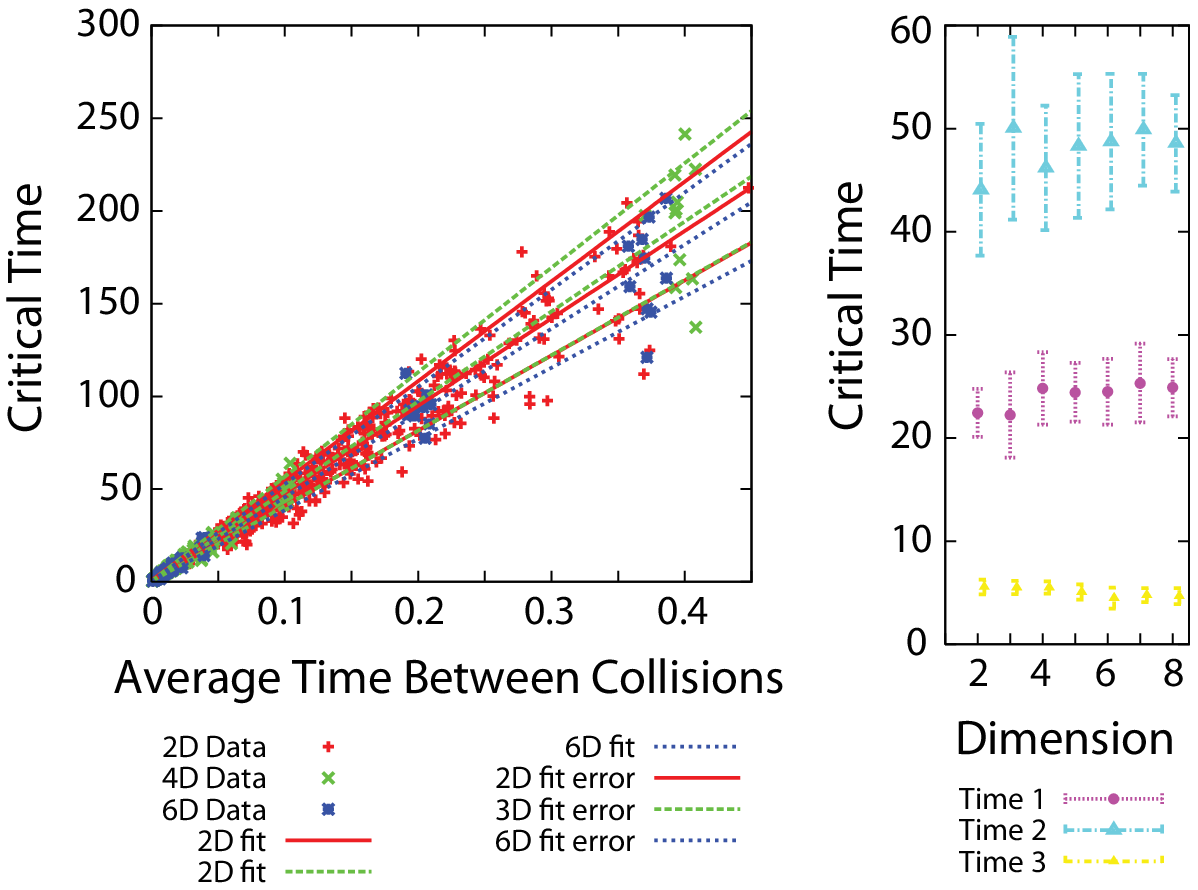}}\\
\subfloat[][]{\includegraphics[trim = 0 0 0 0, scale=0.2]{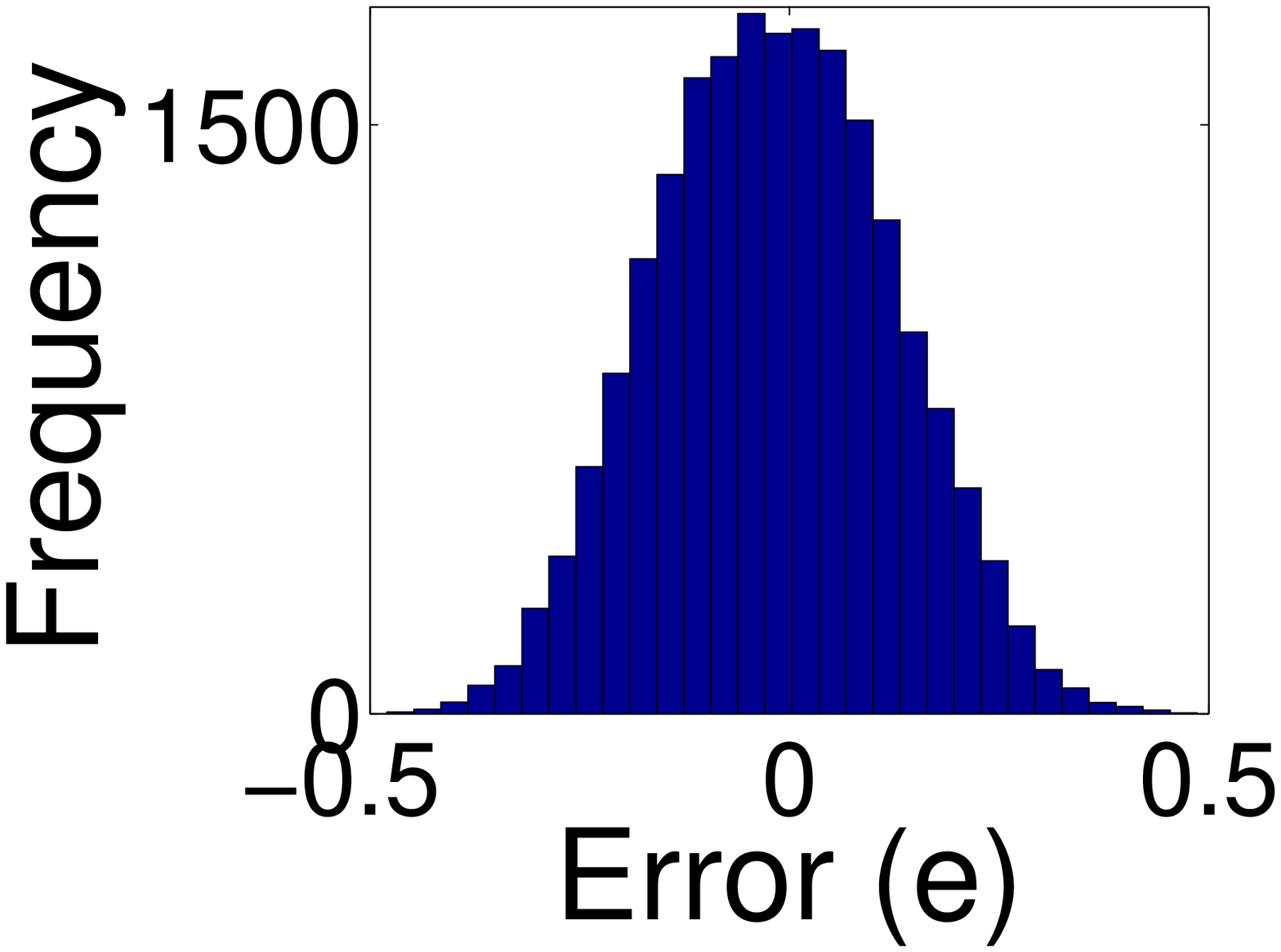}}
\subfloat[][]{\includegraphics[trim = 0 0 0 0, scale=0.2]{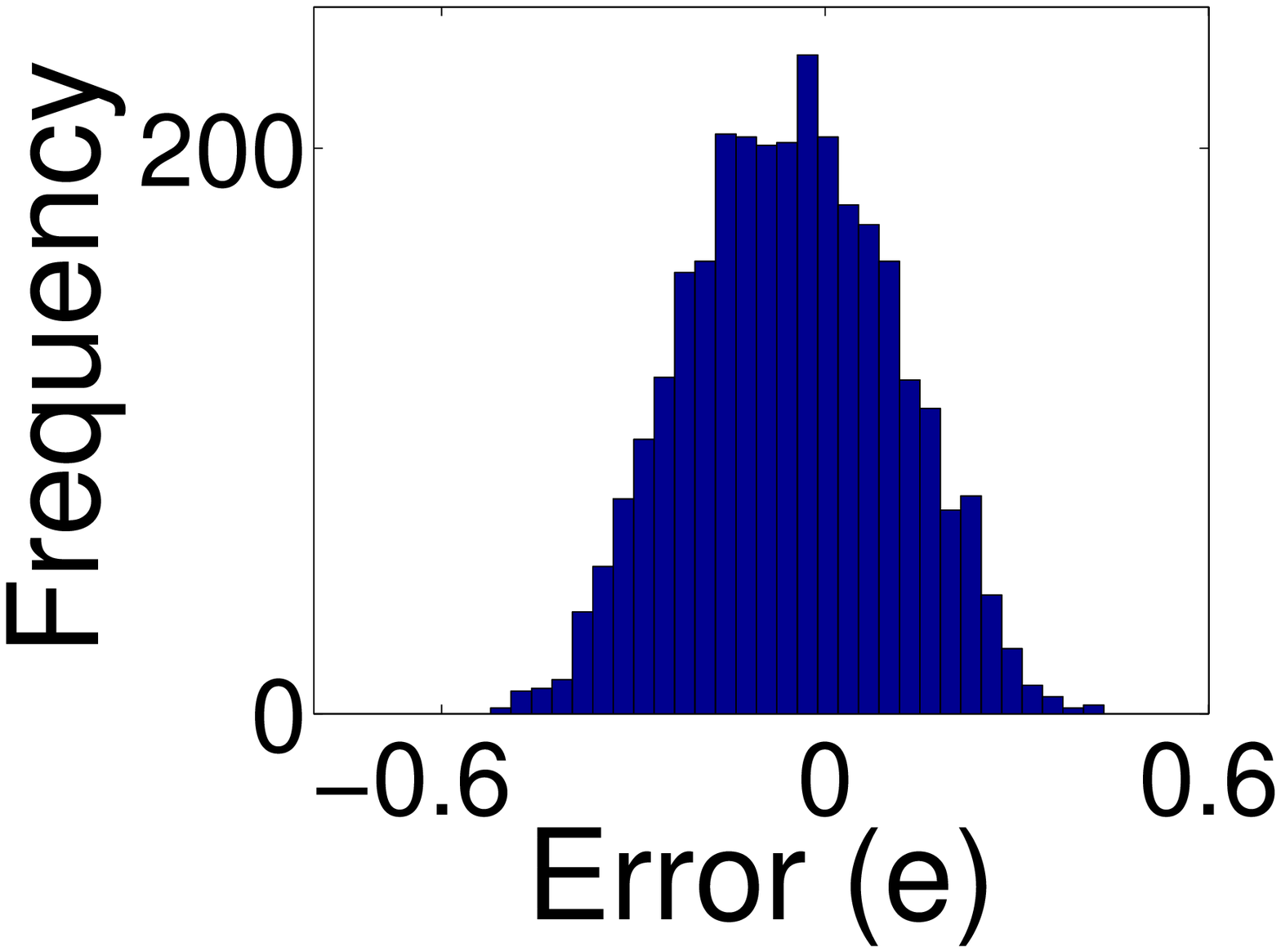}}
\subfloat[][]{\includegraphics[trim = 0 0 0 0, scale=0.2]{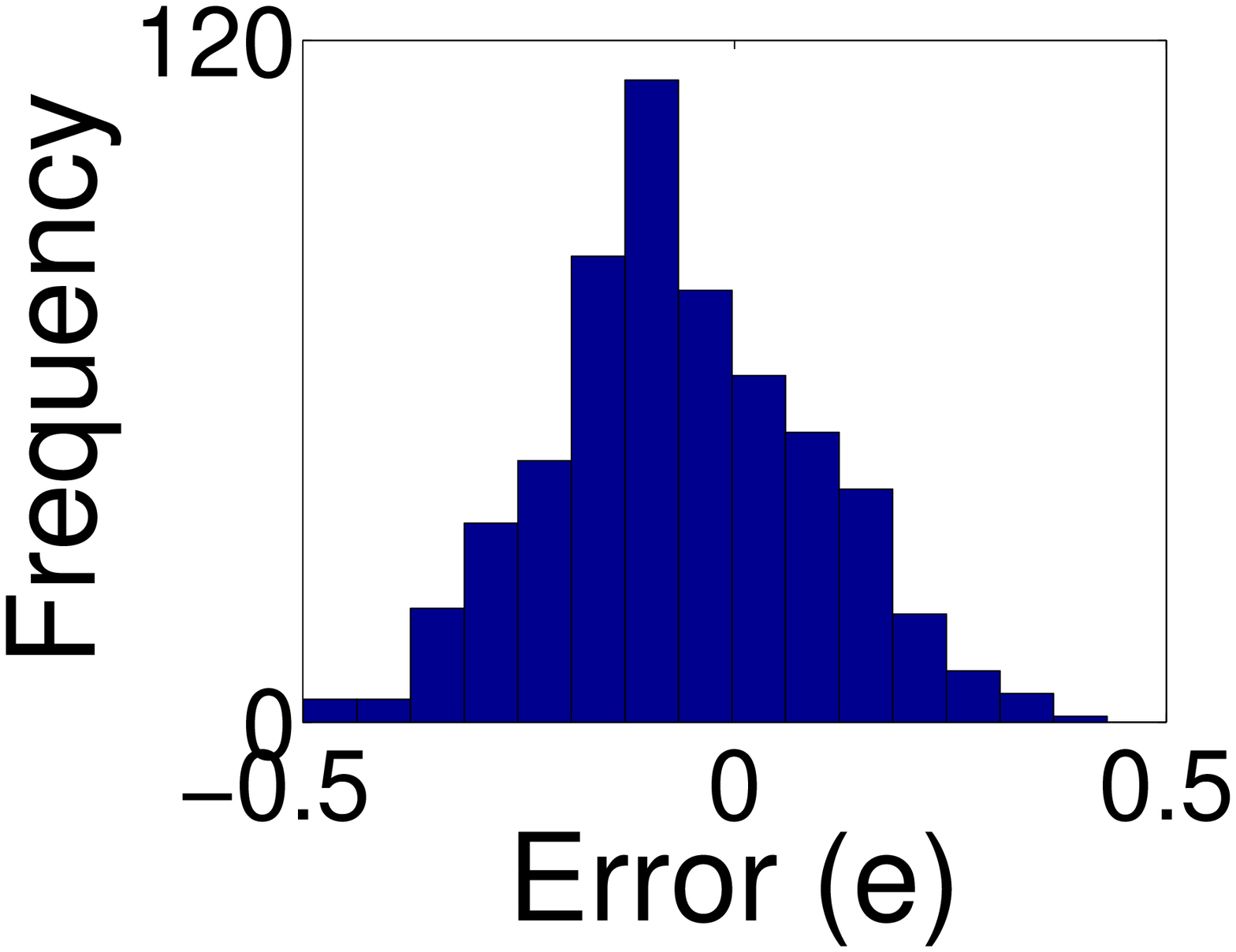}}
\caption{\label{fig:fig4}  (a) Right: Critical times for each dimension for approximately equal times between billiard collisions, $\langle \tau_n \rangle$. For approximately equal $\langle \tau_n \rangle$, critical times are approximately equal in dimensions $n \ge$ 2. Left: for 2D, 4D, and 6D, the critical time, $t_c$, as a function of average time between collisions, $\langle \tau_n \rangle$; each point represents an individual trial for $N$ = 1000 with varying density $0.2<\rho<0.000001$, with $0< \langle \tau_n \rangle <0.45$. The calculated fit for the 2D data is $t_c$ = 0.63 + 471.5$\langle \tau_2 \rangle$, 4D data is $t_c$ = 0.24 + 485.22$\langle \tau_4 \rangle$, and 6D data is $t_c$ = 0.200 + 454.40$\langle \tau_6 \rangle$. This allows us to estimate the critical time through limited statistical sampling of $\langle \tau_n \rangle$. (b)-(d) depict the distribution of errors, $e$. These distributions are approximately gaussian.}
\end{figure*}

These empirical results provide a  predictive tool to estimate the critical time of the system: by estimating $\langle \tau_n \rangle$ from limited statistical sampling, one may predict $t_c$ by using a multiplicative factor. This multiplicative factor is computed from our data in Fig.~\ref{fig:fig4}. Data collected and fit found that for $N$ = 1000, $p$ = 2, the critical time, $t_c$, as a function of average time between collisions, $0< \langle \tau_n \rangle <0.45$, for 2D data is $t_c$ = 0.63 + 471.5$\langle \tau_2 \rangle$, for 4D data is $t_c$ = 0.24 + 485.22$\langle \tau_4 \rangle$, and 6D data is $t_c$ = 0.200 + 454.40$\langle \tau_6 \rangle$.  This relation for  $t_c(\langle \tau_n \rangle)$ must be calculated for each value of $N$.  This gives an expectation value for the critical time, $t_{c,expected}(\langle \tau \rangle)$ based on parameters of the system. The error for actual measurements from the expectation value increases in magnitude proportional to the critical time. The fractional uncertainty for each data point, $e$:

\begin{eqnarray}
e = \frac{t_c,{actual} - t_c,{expected}}{t_c,{expected}}
\end{eqnarray}

\noindent was calculated. The fractional uncertainties of the points form a normal distribution with a standard deviation of $ \sigma_e= 14\% $  for the 2D case, $ \sigma_e= 16\% $  for the 4D case, and $ \sigma_e= 15\% $  for the 6D case. Accounting for this error in the fit relationships, the values fall within each other's standard deviation for points above $t=0.0025$.  We performed additional runs with $1.5< \langle \tau_n \rangle <4.0$ for 2D, 4D, and 6D and found that the previous fits and standard deviations of the fractional uncertainties $\sigma_e$, from the lower $\langle \tau_n \rangle$ data represented well this higher $\langle \tau_n \rangle$ data. In addition, we considered the standard error, comparing the mean critical time for sets of ten runs with equal density to the calculated standard error in time between collisions for that set, $\langle \tau_n \rangle$. As the critical time increases, so does the standard error of $\langle \tau_n \rangle$, as we expect from the broadening data as the critical time increases (Fig.~\ref{fig:fig4}).

Another method for identifying the phase transition, which was proposed in~\cite{gabriel,zaliapin}, analyzes the distribution of clusters $M_{\Delta}^1$ to $M_{\Delta}^{N(t)}$ as function of time $t$.  In Fig.~\ref{fig:fig2} we plot the empirical cluster density distribution, using fifty (50) independent trials with identical initial parameters and compare the magnitude of each cluster ($M^i_{\Delta}(t); 1 < i < N_{\Delta}(t)$) at a time $t$ with the empirical probability density of cluster size.  This empirical density for the cluster size distribution evolves over time according to a power law with exponential taper~\cite{gabriel}: 
\begin{equation}
g_{t}(M) = M^{-\beta}\exp \left(\frac{-M}{\gamma(t)} \right)
\label{eq:19}
\end{equation}

\noindent where the critical exponent is $\beta \approx 2.5$, and $\gamma(t)$ is a time-varying function expressing the tendency for exponential tail in the sample.  As $t$ approaches $t_c$, $\gamma(t)$ diverges and a pure power law emerges.  The power-law fit to our results is shown by a straight line in Figure ~\ref{fig:fig2}. The premonition of catastrophic events could be done by sampling this statistical distribution over time and ascertain the approach to a power law.  This provides an indicator of the imminence of the phase transition.  

\begin{figure}[t]
\centering
\subfloat[][]{\includegraphics[scale=0.65]{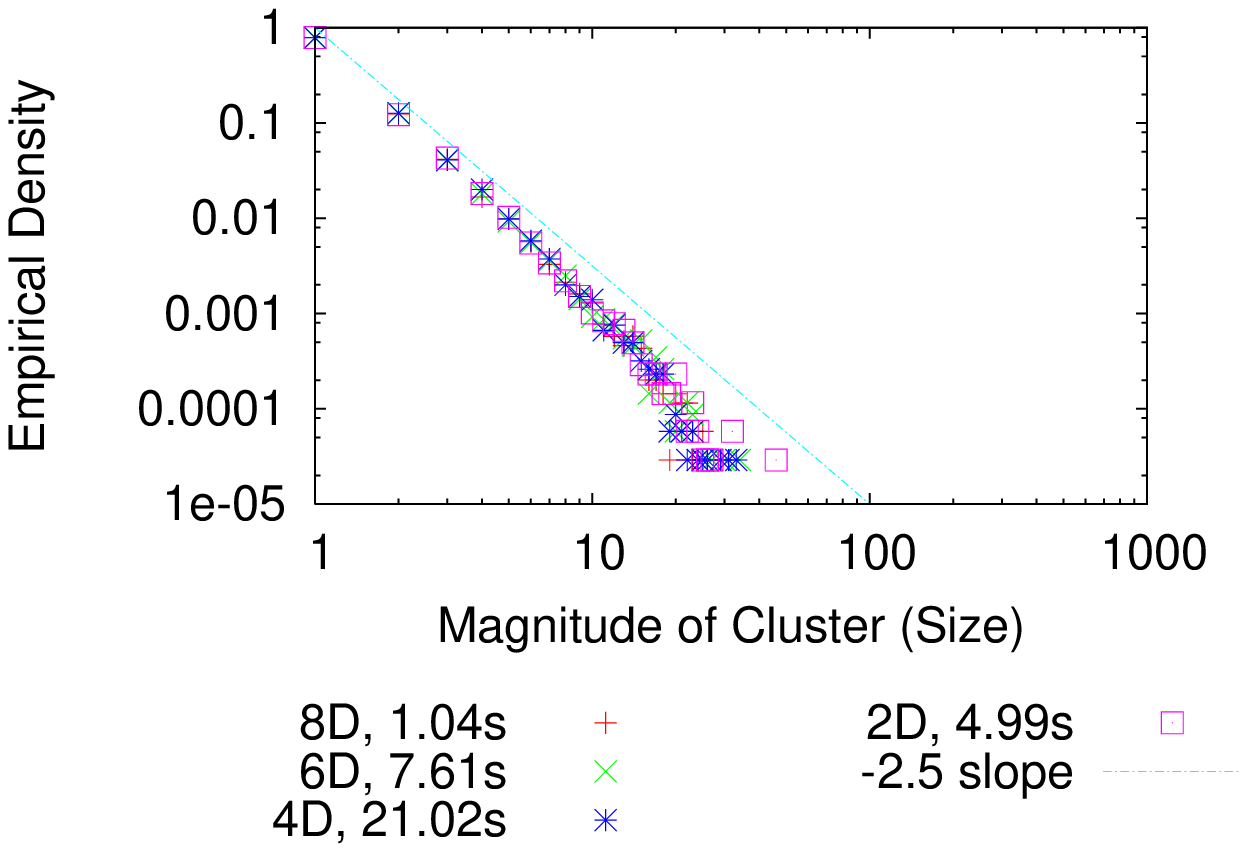}}\\
\subfloat[][]{\includegraphics[scale=0.65]{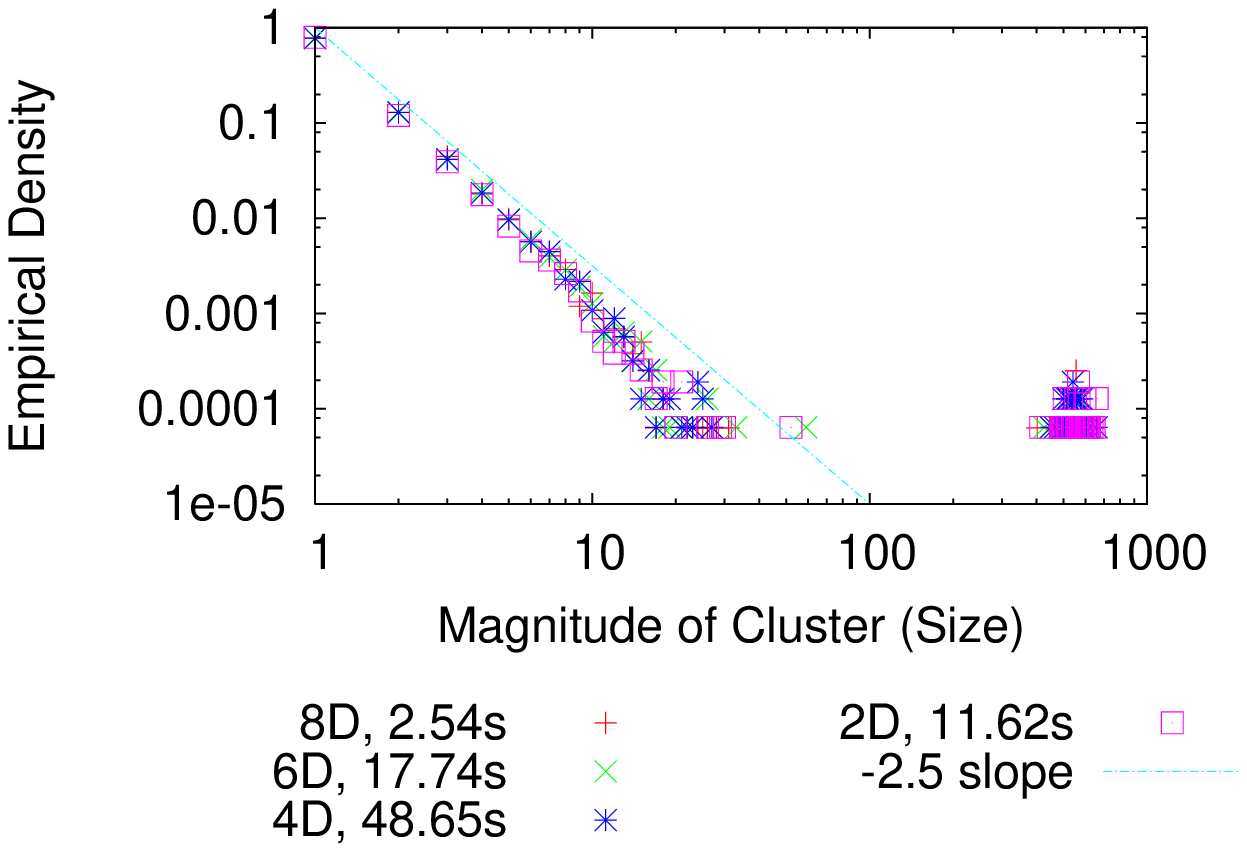}}
\caption{\label{fig:fig5}  Evolution of cluster distribution for $\rho=0.001$, $N=1000$ (a) towards critical time and (b) shortly after the critical time as indicated for dimensions $n$ = 2,4,6,8 as depicted in figure.}
\end{figure}

Nearly identical empirical cluster distributions are found for each dimension $n>2$ at the critical time (Fig.~\ref{fig:fig2}). As for the 2D case of~\cite{gabriel}, varying the density did not result in changes to this distribution.  For each dimension, the cluster distribution evolves according to the power law with exponential tail until the critical time, $t_c$ (see Fig.~\ref{fig:fig5}). The unique distribution at the critical time indicates a critical exponent for the system at the time of phase transition. The critical exponent is consistently $\beta\approx 2.5$ in each dimension studied for elastic collisions.

The usefulness of this premonitory sign in predicting the imminence of a phase transition is determined by the speed at which the power law is approached.  We do this by tracking clusters over time and taking the logarithm of both cluster size and density, and approximate the form of the cluster distribution~(Eq.~\ref{eq:19}) as a low order polynomial in $\log(M)$ at some time $t$:

\begin{equation}\label{eq:21}
\log (g_t(M)) = \kappa_1(t) - \beta\log(M) - \kappa_2(t)\log(M)^2
\end{equation}

\noindent The time-evolution of the cluster distribution is recorded according to Eq.~(\ref{eq:21}), fixing $\beta = 2.5$, and fitting the parameters $\kappa_1(t), \kappa_2(t)$ for each $t$. A pure power law emerges at the critical time, as $\kappa_2(t) \rightarrow$ 0.

\begin{figure}[t]
\includegraphics[trim = 0 0 0 0, scale=0.7]{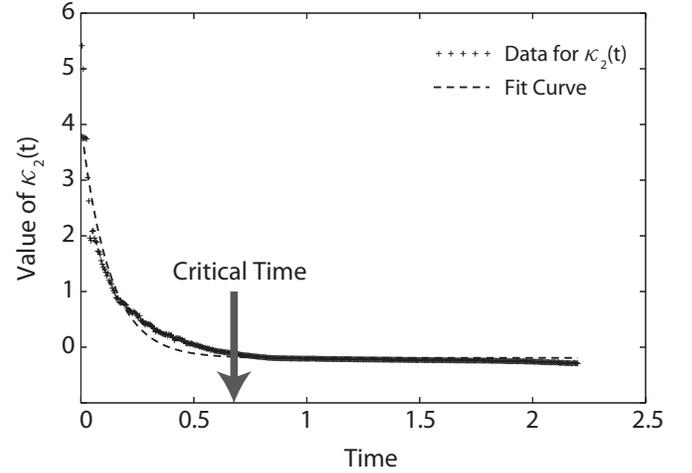}
\caption{\label{fig:fig6}  Evolution of $\kappa_2(t)$ averaged for 50 samples, each in 4D, $N$ = 1000, $\rho = 0.1$.  The average critical time for the 50 samples was  0.707; the standard deviation for the critical times was 0.10.}
\end{figure}

Cluster density distributions $g_t(M)$ were created by running simulations fifty times for various parameters as reported in Table~\ref{tab:1}. Results for the time-evolution of the coefficient $\kappa_2(t)$ are presented in Fig.~\ref{fig:fig6}. The parameter is seen to decay approximately exponentially.  The $1/e$ point, denoted $t^*$, can be used as a predictive criterion. Fitting the $\kappa_2$ values using $\kappa_2 = A_1 \cdot \exp[-(t-A_2)/t^*] + B_1 $, we find values for $t^*$ tabulated in Table~\ref{tab:1}. The data fits in Table~\ref{tab:1} show that a pure power law fit is first approached approximately two to four standard deviations, $\sigma_{t_c}$, prior to the mean, $\langle t_c \rangle$, for fifty runs.

\begin{table}[ht]
\caption{Fitting results for the function (21) parameters. $n$ is the dimension (an integer), $\rho$ is the density, $N$ is the number of billiards, $\langle t_c \rangle$ is the average critical time, $\sigma_{t_c}$ is the standard deviation from the critical time, $\kappa_2(\langle {t_c} \rangle)$ is the fit parameter at the average critical time, $t$ at $ \kappa_2(t)\approx0$ is the time when the pure power law occurs. \label{tab:1} }
\centering
\begin{tabular}{c c c c c c c c}
\hline\hline 
$n$ & $\rho$ & $N$ & $\langle t_c \rangle$ & $\sigma_{t_c}$ & $\kappa_2(\langle t_c \rangle)$  & $t^*$ & $t$ at $ \kappa_2(t)\approx0$\\ [0.5ex] 
\hline 
2 & 0.1 & 1000 & 1.271 & 0.1755 & -0.1324 & 0.16601 & 0.84\\
2 & 0.01 & 1000 & 3.1584 & 0.3167 & -0.1525 & 0.410102 & 2.3\\
2 & 0.001 & 1000 & 7.8147 & 1.0354 & -0.1312 & 1.3766 & 6.0\\
2 & 0.01 & 2500 & 3.258 & 0.2491 & -0.153 & 0.351295 & 2.47\\
4 & 0.1 & 1000 & 0.70751 & 0.09624 & -0.1339 & 0.126561 & 0.54\\
4 & 0.01 & 1000 & 5.6819 & 0.8655 & -0.1287 & 0.74666 & 4.18\\
4 & 0.001 & 1000 & 32.208 & 3.931 & -0.1146 & 4.77281 & 26.6\\
4 & 0.01 & 2500 & 7.5752 & 0.6641 & -0.1306 & 0.680786 & 6.175\\
6 & 0.01 & 2500 & 1.9977 & 0.2280 & -0.125 & 0.266022 & 1.54\\
6 & 0.01 & 1000 & 1.3255 & 0.2133 & -0.1147 & 0.203752 & 1.02\\
6 & 0.001 & 1000 & 11.708 & 1.516 & -0.1242 & 1.86351 & 9.6\\ [1ex]
\hline 
\end{tabular}
\label{table:nonlin}
\end{table}

\begin{figure}[t]
\includegraphics[scale=0.7]{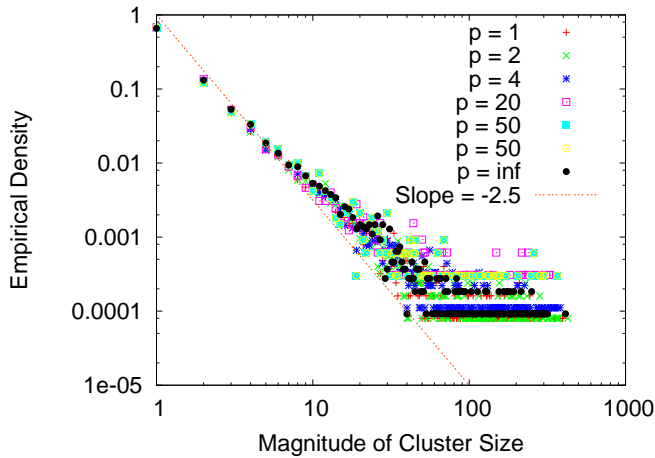}
\caption{\label{fig:fig7}  Empirical cluster distribution for varying L$^p$ -norm spaces as indicated for $N=1000$, $\rho=10^{-1}$ for 50 runs.}
\end{figure}


Future research may consider the possible existence for an upper critical dimension where the $\beta=5/2$ is no longer valid. Previous work has found for a particular system of coalescing clusters that there is an exact solution of the Smoluchowski equation valid for $n>n_c=2$ ~\cite{kang}. It is possible that the reason for the observed universality for $n>2$ could be that these models all exist above a certain critical dimension and the results we see are independent of dimension.

\subsection{L$^p$ Norm Results}

We now look at dynamical transitions in 2D, but using the L$^p$ norm metric rather than the Euclidean distance.  There is an approximately equivalent power law distribution at the critical time, as shown in Fig.~\ref{fig:fig7} for $N=1000$ and for varying $p$, in the range: $1\le p \le \infty$ at the same density, $\rho$.  When L$^p$ norms are used to check for hard sphere overlap, some amount of interpenetration of particles is permitted, as can be seen by the fact that an L$^p$-norm can exaggerate the importance of dominant vector components regardless of the contribution from remaining components.  Models of granular media based on L$^p$ norms have been used to study collisions between non-spherical particles~\cite{gamba}. 

\subsection{The Coalescence Case} 
 
\begin{figure}[t]
\includegraphics[scale=0.7]{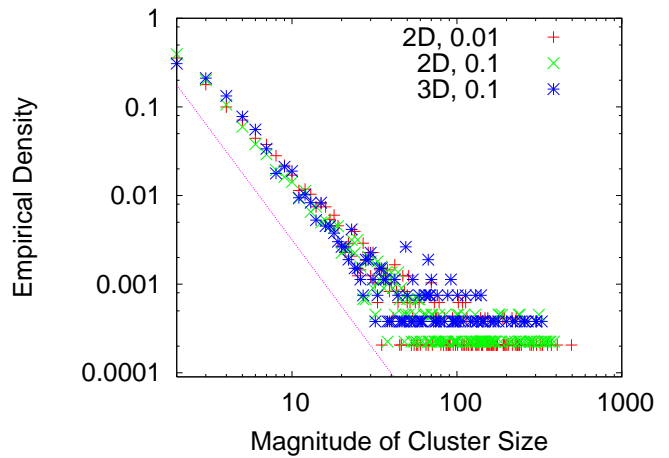}
\caption{\label{fig:fig9} Empirical cluster density at critical time, $t_c$ for $E_a=0$, with densities and dimensions as indicated. }
\end{figure}

Coalescence dynamics are relevant to many physical processes such as nanoparticle and colloidal growth.  We consider only dynamical systems in which the total energy is conserved. Our investigation here focuses on whether the trends observed in the Sinai billiard case, namely the behavior of the empirical cluster distribution, extend to the coalescence billiard case.

\begin{figure}[t]
\centering
\includegraphics[scale=0.7]{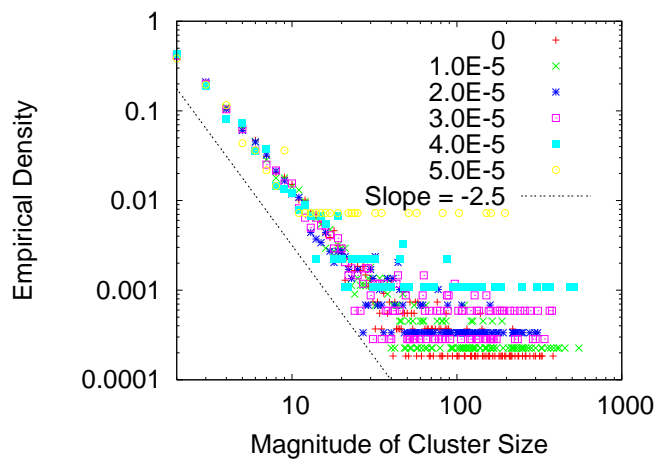}
\caption{\label{fig:fig10} Empirical cluster density at critical time, for 2-dimensional, $N$ = 1000, $\rho=$0.1, at $t_c$ for the case of varying percentage of effective collisions forming coalesced billiard, with varying $E_a$ as indicated. }
\end{figure}

The empirical cluster mass distribution at $t_c$ has a power law at $t_c$, as seen in Fig.~\ref{fig:fig9} for the model where all collisions result in coalescence.  This is similar to what we observed in the elastic billiards case.  For the coalescing cluster distributions, we only considered clusters which have been involved in at least one coalescing collision (of mass $M\ge 2.0/N$).  Billiards which have not coalesced are not considered, as they do not satisfy the $\Delta$-neighbor relationship with any other billiards. The coalescing cluster distribution plot compares the magnitude of each cluster [$2 < i < N_{\Delta, S}(t)$] against its empirical probability.

As $E_a$ increases, $t_c$ also increases due to the decreasing fraction of successful coalescing collisions. For higher $E_a$, fewer small clusters are formed and large clusters dominate throughout the trial. Figure~\ref{fig:fig10} shows these distributions over multiple runs for different values of $E_a$. The critical exponent fit is consistent across all models, indicating a large degree of universality in the system. One possible reason for this universality is that all models are at the critical dimension.  Such a behavior has been observed by Kang in the case of the Smoluchowski equation and $n=2$~\cite{kang} in the elastic case. Our work could be further extended to investigate the form of the kernel for Smoluchowski equation for this model.

The analysis of dynamical phase transitions is relevant to coalescing clusters.  Coalescent events have been studied recently in nanoparticle growth trajectories ~\cite{zheng}. This study may be useful in modeling such phenomena.

\section{Conclusion} 

We have extended the analysis of dynamical phase transitions to higher dimensions, densities, and norms.  We have also considered two cases of collisions: elastic collisions and coalescing billiards, and for the elastic case the analysis included the effects of an L$^p$ norm.  We have found universality in the form of a power law for the probability distribution of cluster sizes, with the same critical exponent describing these  systems.  In the non-coalescing case, the expectation value of the critical time was shown to be determined mainly by the average time between collisions.  These observations can be used to predict the onset of criticality and could be used for applications in chemistry such as gas dynamics or polymerization.

\bibliographystyle{unsrt}
\bibliography{refs}

\end{document}